\documentstyle[12pt,epsfig]{article}
\begin{document}
\baselineskip 3.9ex
\def\l#1{\label{eq:#1}}
\def\eqn#1{(~\ref{eq:#1}~)}
\def\no{\nonumber} 
\def\av#1{{\langle  #1 \rangle}}
\title{ Directed Branched Polymers near an Attractive Line\\
\vspace{0.2cm}
 Sumedha\thanks{sumedha@theory.tifr.res.in}\\
\vspace{0.2cm} 
Department Of Theoretical Physics\\ Tata
Institute Of Fundamental Research\\
Homi Bhabha Road, Mumbai 400005\\
India}
\date{28 January 2004}
\maketitle
\begin{abstract}
We  study  the  adsorption-desorption  phase  transition  of  directed
branched polymer in $d+1$ dimensions in contact with a line by mapping
it to a $d$ dimensional hard core lattice gas at negative activity. We
solve the model exactly in $1+1$ dimensions, and calculate the crossover
exponent related to fraction of monomers adsorbed at
the critical point of surface transition, and  we also determine the  density  profile
of  the  polymer  in different phases.  We  also obtain the value of
crossover exponent in $2+1$ dimensions  and give  the scaling function
of the sticking fraction for $1+1$  and $2+1$ dimensional directed
branched polymer.
\end{abstract} 
  
 Linear and branched polymers,  near an attractive surface, undergo an
adsorption-desorption transition, which  has important applications in
areas  ranging  from  technology  such as  in  lubrication,  adhesion,
surface  protection to  biology  \cite{eisenriegler,vanderzande}.  For
example,  adsorbed  polymers  are  used  for  surface-modification  of
medical  implants \cite{norde}.  There  have been  several theoretical
studies   of  the   behavior   of   a  polymer near a
surface\cite{vanderzande,bell,stella,privman,singh,bouchaud}. Especially,
effect  of surface  for  idealized polymer  (with no  self-exclusion),
modeled by random walks has  been studied extensively.  There are many
exact results known for Gaussian random walks in presence of a surface
\cite{rubin,satya}. In comparison,  linear polymer with self exclusion
and branched polymer  are less well studied. For  a self-avoiding walk
(SAW) in  the vicinity of a  surface the exact  critical exponents are
known  from conformal field theory \cite{duplantier}. Directed polymer
chain adsorption, modeled by a  directed SAW is one of the few
solvable   models  of   surface  effects   in  2   and   3  dimensions
\cite{privman}. For directed walks, self exclusion is automatic, and
nontrivial effects of excluded volume interaction are not seen. For
branched polymers (modeled by lattice animals), a relation between the
exponent characterizing the number of animals, in presence of
surface, and  in the bulk is known from a
simple argument given by  De'Bell et. al  \cite{debell}. In
this paper  we solve directed branched  polymer (DBP) in $2$
dimensions  and $3$ dimensions in presence of a $1d$  line
exactly. Introducing a preferred direction makes the system
analytically  more tractable. Similar results for $2$ dimensions have
also been obtained by Rensburg et. al. \cite{rensburg} independently.

The  enumeration of directed  site animals  in $d  + 1$  dimensions is
related  to hard-core  lattice gas  (HCLG) at  negative  activity with
repulsive interactions in $d$ dimensions and the Yang-Lee edge problem
in  $d$ dimensions  \cite{cardy,dhar1,lai,brydges}. In  this  paper we
give the mapping  of a $d+1$ dimensional directed  branched polymer in
presence  of  a  line  to   a  $d$  dimensional  HCLG  with  repulsive
interactions. 

	The plan of the paper is as follows. In section 1 we will
define the model of directed branched polymer and the quantities of
interest. Using the above mentioned correspondence we give the mapping
of a $d+1$ dimensional directed branched polymer in presence of a line
to a $d$ dimensional HCLG with repulsive interactions in Section 2. In
Section 3, for $1+1$ $d$ DBP in presence of a $1d$ penetrable line  we
solve the model exactly. Section 4 deals with the DBP in $1+1$ $d$ in
presence of a impenetrable line. For DBP in $1+1$ dimensions we show
that the behavior at the transition point for penetrable and
impenetrable wall is the same, and not just the crossover exponent but
even the density profile is the same. This implies that for $1+1$
dimensions, for impenetrable surface, at the phase transition point
the decrease in entropy is exactly compensated by the increase in
internal energy. This seems to be a special property of polymers in
$2$ dimensions. Even for linear polymers the exponent for both cases
is the same and hence it is believed that for a linear polymer in $2$
dimensions in presence of a impenetrable surface the phase transition
point correspond to point where surface effects vanish completely and
system behaves like bulk \cite{bouchaud}. Here we are able to show it
explicitly for directed branched polymers. In Section 5 using Baxter's
solution of hard hexagon gas we study $2+1$ dimensional DBP in
presence of line and calculate the crossover exponent and sticking
fraction for the directed branched polymer exactly. The scaling
function of sticking fraction is a function of two intensive
thermodynamic variables. We have derived its exact form in $1+1$
dimension and $2+1$ dimension. There are very few such exact
nontrivial scaling functions of more than one thermodynamic variable
known \cite{cardy1}. We also get the large $w$ expansion of the
sticking fraction as a power series in $1/\sqrt{w}$.

\section{The Model}

A directed branched polymer or a directed animal on a lattice, rooted
 at the origin is a connected cluster such that any site of the animal
 can be reached from the root by a walk which never goes opposite to
 the preferred direction. For example, on a square lattice drawn
 tilted at $45\%$ in Fig. 1, a directed site animal or a directed
 branched polymer $\mathcal{A}$ rooted at the origin is a set of
 occupied sites including origin, such that for each occupied site
 $(x,t)$ other than the origin, at least one of the two sites
 $(x-1,t-1)$ and $(x+1,t-1)$ is also occupied. The number of sites in
 ${\mathcal{A}}$ will be denoted by $s = |\mathcal{A}|$. We define
 $n(x|{\mathcal{A}})$ as the number of sites of $\mathcal{A}$ having
 the transverse coordinate $x$. We study the DBP in presence of $1d$
 line parallel to the preferred direction. This is positioned along
 the main diagonal of the lattice (Fig. 1). We will consider only
 polymers rooted at the surface in this paper.

 We assign a fugacity $y$ to all allowed
  sites of the cluster. Further, if we associate an additional energy $-E$
   with each site on the surface, each site on surface will have an
   additional weight and the fugacity of sites about the diagonal,
  	       denoted by $y_0$ is equal to $w y$ where
\begin{equation}
			w = \mbox{exp} (E/kT)  \label{act}
\end{equation} 
	 Hence $w > 1$ would correspond to an attractive surface.

 We define $A(w,y)$, the grand partition function of the polymer as

\begin{equation}
      A(w,y) = \sum_{\mathcal{A}} y^{|{\mathcal{A}}|} w^{n_0} =
		   \sum_{s=1}^{\infty}  A_s(w) y^s
\end{equation}
where $n_0 = n(0|{\mathcal{A}})$ and $A_s(w)$ is the partition
function of the polymer made of exactly $s$ monomers.

For $w=1$, we get the statistics of equally weighted animals and
 $A_s(1)$ is the number of distinct directed animals having $s$ sites
 with given boundary conditions. For large $s$, $A_s(w)$ varies as
 ${\lambda}^s s^{\theta}$, where $\theta$ is known as entropic
 critical exponent. Similarly, the transverse size of the polymer for
 large $s$ scales as $s^{\nu}$, where $\nu$ is the exponent which
 defines the transverse length scale of the polymer. These exponents
 take different values in desorbed , adsorbed and in the critical
 regions. We will use subscripts $de$, $c$ and $ad$ to represent
 critical exponents and other quantities in desorbed , critical and
 adsorbed phases of the polymer.

And the free energy per monomer of the polymer in thermodynamic limit
			     is given by
\begin{equation}
 F(T) = \lim_{s \rightarrow \infty} -\frac{k_B T}{s} log A_s(w) \equiv
		      k_B T log({y_{\infty}(w)})
\label{free}
\end{equation}
where $y_{\infty}(w)$ is the value of fugacity at which $A(w,y)$ has a
singularity for a given value of $w$.

\begin{figure}
\begin{center}
		  \epsfig{figure=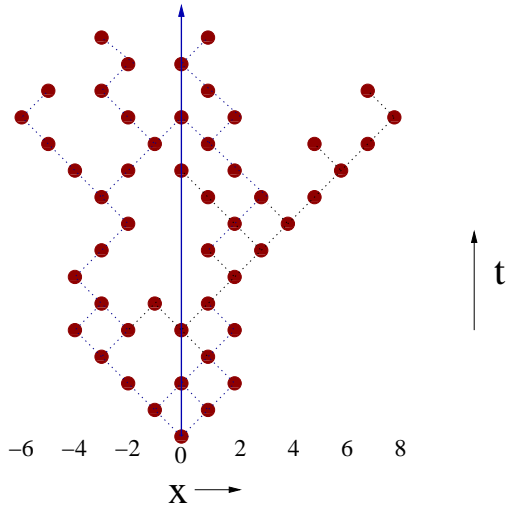,width=7cm}
\end{center}
 \caption{A Directed Branched Polymer of size 50, rooted on the surface.}
\end{figure}

Let $\phi({\textbf{x}},{s})$ be the  value of
       $n({\textbf{x}},{\mathcal{A}})$ averaged over all configurations
       $\mathcal{A}$ of size $s$. We define a generating function
       $\Psi(\textbf{x};w,y)$ as

\begin{equation} 
\Psi(\textbf{x};w,y) = \sum_{\mathcal{A}} n({\textbf{x}}|{\mathcal{A}})
		     w^{n_0} y^{|{\mathcal{A}}|} \equiv \sum_{s}
		     \phi(\textbf{x},{s}) A_s(w)~ y^s
\end{equation}
   
There is a critical value $w_c$ of wall activity such that for $w >
       w_c$, $\phi(0,{s})$ is proportional to $s$ for large $s$ and
       the transverse size is finite $(\nu_{ad}=0)$. This is the
       adsorbed phase, in which monomers tend to stick to the
       surface. $w < w_c$ corresponds to the desorbed phase of the
       polymer in which only a finite number of monomers stick to the
       surface. At $w=w_c$, the critical point of the surface
       transition, the number of adsorbed monomers as function of
       polymer size in large $s$ limit have a behavior given by

\begin{equation}
		     \phi_c(0,s) \sim s^{\alpha};
\end{equation}
where $\alpha$ is known as the crossover exponent of the surface
  transition.

In the $s \rightarrow \infty$ limit, the fraction of monomers adsorbed
  is like order parameter of the surface phase transition. In the
  constant fugacity ensemble $A(w,y)$ is the partition function with
  fixed $w$ and $y$ and hence the average polymer size would be given
  by

\begin{equation}
   \langle s(y,w) \rangle = \frac{\sum s y^s w^{n_0}}{\sum y^s
	     w^{n_0}} \equiv \frac{\partial \mbox{ln} A(w,y)}{\partial
	     \mbox{ln} y}
\end{equation}

   Similarly, the average number of monomers at the surface would be

\begin{equation}
 \langle n_0(y,w)\rangle = \frac{\sum n_0 y^s w^{n_0}}{\sum y^s
	     w^{n_0}} \equiv \frac{\partial \mbox{ln} A(w,y)}{\partial
	     \mbox{ln} w}
\end{equation}

The sticking fraction defined as the fraction of polymer segments at
the surface, represented by $C_{st}(w,y)$, would be given by

\begin{equation}
C_{st}(w,y) = \frac{ \langle n_0(y,w)\rangle}{ \langle s(y,w)\rangle}
\end{equation}

In the infinite polymer limit, if we represent the value of fugacity
at which $\langle s(y,w) \rangle$ diverges by $y_{\infty}(w)$ for a
given $w$, then the sticking fraction is only a function of the wall
activity $w$ and is given by

\begin{equation}
   C_{st}(w) = - \frac{d \mbox{ln}y_{\infty}(w)}{d\mbox{ln}w}
\end{equation}

This is the order parameter of the surface phase transition and is
zero for $w \leq w_c$, where $w_c$ is the surface phase transition
point.

In general, in the large polymer limit, near critical value of $w$, as
$ w \rightarrow w_c^{+}$, $C_{st}(w,y)$ is expected to have scaling
form

\begin{equation}
   C_{st}(w,y) = \epsilon^{1-\alpha} h((w-w_c) \epsilon^{-\alpha})
   \label{stscale}
\end{equation}
where $\epsilon = 1- y/y_{\infty}(w)$. The scaling function $h(u)$
where $u=(w-w_c) \epsilon^{-\alpha}$, is a function of $w$ and $y$
which are both intensive thermodynamic variables. As $u \rightarrow
\infty$, $h(u) \sim u^{(1-\alpha)/\alpha}$.

\section{General Results}

The directed site animal enumeration (DSAE)  problem in
   $d+1$-dimensions is related to time development of thermal
   relaxation of a hard core lattice gas (HCLG) with nearest neighbor
   exclusion on $d$ dimensional lattice \cite{dhar1}. In \cite{sumedha1}, we
   have shown that this correspondence relates the density at a site
   $i$ in steady state to sum of weights of all animals rooted at $i$,
   the grand partition function of the animal. Also, the average
   number of sites at a given transverse distance $\textbf{x}$ from
   the origin for a $d+1$ dimensional directed animal is related to
   the density-density correlation function of the lattice gas in $d$
   dimensions.

Specifically, if on a $d+1$ dimensional body-centered hyper-cubic
lattice we define weight of an animal $\mathcal{A}$ as the product of
weights of all occupied sites, with weight corresponding to a site
with $\textbf{x}$ coordinate $i$ being $y_i$, then the DSAE problem on
this $d+1$ dimensional lattice gets related to time development of
HCLG with nearest neighbor exclusion on a $d$ dimensional
body-centered hyper-cubic lattice with the rates which satisfy
detailed balance condition corresponding to the Hamiltonian

\begin{equation}
H = +\infty \sum_{<ij>} {n_i n_j} -  \sum_{i} (\mbox{ln}z_i) { n_i}   
\end{equation}
where $z_i = -y_i/(1+y_i)$ and the animal number generating function
is just the negative of density of HCLG with change of variables from
$z$ to $y$. Here we have used the convention that if $\sum_{<ij>} {n_i
n_j}=0$ then the corresponding term in the Hamiltonian is zero. The
configurations with any pair of occupied nearest neighbor have
infinite energy and do not contribute to the partition function.

The partition function is linear in all $z_i's$. The linearity of the
  partition function in $z_i's$ implies that in case when the activity
  about $\textbf{x} = 0$ is different from that in the rest of the
  sample, i.e if we let the activity about $\textbf{x} = 0$ be $z_0$
  and activity in rest of space be $z$, then the partition function of
  the HCLG can be written as

\begin{equation}
Z(z_0,z) = A(z)+z_0 B(z)
\end{equation}
where $A(z)$ and $B(z)$ are polynomials in $z$. If $\rho$  represents
the density of HCLG when the activity about each site is the same,
then the density of HCLG about the origin in the present case
$\rho_0(z_0,z)$ can be written in terms of $\rho$ as

\begin{equation}
\rho_0(z_0,z) = \frac{z_0 \rho}{\rho z_0+z (1-\rho)}
\end{equation}

Same observation has been made by Cardy in
\cite{cardy2}. Correspondingly, since $A(w,y)$ is just the negative of
$\rho_0(z_0,z)$ with $z_0 = -wy/(1+wy)$ and $z=-y/(1+y)$, we can
express $A(w,y)$ in terms of $A(1,y)$ and this is given by

\begin{equation} 
   A(w,y) = \frac{w(1+y)A(1,y)}{(1+wy)+A(1,y)(1-w)} \label{density}
\end{equation}

  Moreover, the density-density correlation function of HCLG
   $G(\textbf{x};w,z)$ with $w\neq 1$ can be expressed in terms of
   density density correlation function when $w=1$. We find that the
   density density correlation function is related to
   $\Psi(\textbf{x};w,y)$ on a hyper-cubic lattice as follows
\begin{equation}
\Psi(\textbf{x};w,y) = -\frac{1}{1+y}
G\left(\textbf{x};w,z=\frac{-y}{1+y}\right)    \label{rel}
\end{equation}
from this we get
\begin{equation}
\frac{\Psi(\textbf{x};w,y)}{\Psi(\textbf{x};1,y)} =
 \frac{w(1+y)[1+wy-(1-A(1,y))(1-w)]}{[1+wy+A(1,y)(1-w)]^2} \label{corr}
\end{equation}

Since $\rho$ is the density of the HCLG, then as discussed in
   \cite{sumedha1}, for $\textbf{x} = 0$, the density density
   correlation of HCLG is always equal to $\rho (\rho -1)$ for any $d$
   dimensional case and hence $\Psi(0;1,y)$ can be completely
   expressed in terms of $A(1,y)$. Hence we get,

\begin{equation}
  \Psi(0;w,y) = \frac{w(1+y)A(1,y)(1+A(1,y))}{[1+wy+A(1,y)(1-w)]^2}
			    \label{stick}
\end{equation}

Eq. (\ref{density}-\ref{stick}) hold for all dimensions. Hence, in
  presence of $1d$ surface, a DBP in $d+1$ dimensions rooted on the
  surface can be studied using the mapping to HCLG. Moreover the
  generating functions $A(w,y)$ and $\Psi(0;w,y)$ can be completely
  expressed in terms of animal number generating function when wall is
  neutral i.e, in terms of $A(1,y)$. We will use these results in rest
  of the paper to study the surface effects for DBP in $2$ and $3$
  dimensions.

In the adsorbed regime the number of monomers in direct contact with
     the wall is proportional to $s$ and $\nu_{ad} = 0$. This implies
     that the scaling form of $\phi(\textbf{x},s)$ in the adsorbed
     regime would be

\begin{equation}
       \phi_{ad}({\bf{x}},s) \sim \frac{s}{\xi^d}g(|{\bf{x}}|/\xi)
       \label{ads}
\end{equation}
where $\xi = (w-w_c)^{\tilde{\nu}}$ is the chracterstic length scale
in the system. Since we are away from the critical regime, $\xi$ is
well behaved and never diverges for finite $w$. Also $\xi$ is
independent of the size $s$ of the polymers. The normalization of
scaling function $g(r)$ is chosen such that
\begin{equation}
\int_{-\infty}^{\infty} d^d \textbf{x} g(|\textbf{x}|) = 1
\end{equation}

$A_s(w)\phi(\textbf{x},s)$ is the coefficient of $y^s$ in the
expansion of $\Psi({\bf{x}};w,y)$. In the adsorbed regime, $A_s(w)
\sim (y_{\infty}(w))^{-s}$ for large $s$ and behavior of
$\phi({\bf{x}},s)$ is given by Eq. (\ref{ads}), hence
$\Psi({\bf{x}};w,y)$ will have a scaling form
\begin{equation}
	\Psi({\bf{x}};w,y) \sim \frac{\epsilon^{-2}}{\xi^d}
	g(|{\bf{x}}|/\xi)
\end{equation}
where $\epsilon =  1-y/{y_{\infty}(w)}$.

Since the scaling function $g(|{\bf{x}}|/\xi)$ has no $y$ dependence,
hence the scaling function of $G({\bf{x}};w,z)$ would also be just
$g(|{\bf{x}}|/\xi)$ for $w>w_c$.

\section{Two dimensional Directed Branched Polymer in presence of 1-d penetrable surface}

  For a penetrable surface, since the configurations spanning through
  the surface are allowed, there is no loss of entropy per monomer to
  take into account (Fig 1). Hence, $w=1$ corresponds to a zero gain
  in free energy per monomer of the surface. We find that the value of
  $y$ at which  $A(1,y)$ diverges, the only value of $w$ which makes
  $A(w,y)$ also divergent is $w=1$. This implies that $w_c =1$ for a
  DBP in any dimension in presence of a $1d$ line as long as $A(1,y)$
  gets singular at finite value of $y$. Then the polymer has bulk
  behavior at the critical point. At $w=1$, i.e for directed branched
  polymer in bulk, we have shown in an earlier paper \cite{sumedha1} by
  scaling arguments and dimensional analysis $\phi({\bf{x}},s)$ has a
  scaling form

\begin{equation}
	       \phi_c(\textbf{x},s) \sim {s}^{1-d\nu_c}
		  f({|\textbf{x}|}{\epsilon}^{-\nu_c})
\end{equation}

       This implies $\phi_c(0,s) \sim s^{1- d \nu_{c}}$, and the
			       crossover exponent $\alpha $ is exactly
			       given by

\begin{equation}
	 \alpha = 1 - d \nu_{c} \equiv 1-\theta   \label{alp}
\end{equation}
where $\nu_c$  is the transverse correlation exponent of a $d+1$
dimensional DA in bulk, which is equal to the correlation length
exponent for a $d$ dimensional HCLG with nearest neighbor exclusion.

As we go to higher dimensions even though entropy loss and energy gain
balances each other at $w=1$, the polymer might start binding to a
line only at wall activity greater than $1$. For directed branched
polymers, when $A(1,y)$ has no divergence, $w=1$ is not the critical
point of the surface transition. Instead it is given by

\begin{equation}
w_c= \frac{1+1/A(1,y_c)}{1-y_c/A(1,y_c)} \label{pw}
\end{equation}
where $y_c$ is the large polymer limit fugacity value of the polymer
with neutral wall i.e, when $w=1$.

As an example, on a Bethe lattice with co-ordination number $3$ the
function $A(1,y)$ is
\begin{equation}
A_B(1,y) = \frac{1-\sqrt{1-4y}}{2y}
\end{equation}
and $y_c=1/4$. At $y=1/4$ the function $A_B(1,y)=2$, and substituting in Eq.\ref{pw} we get $w_c =12/7$, which is greater than
$1$.

 The $1+1$ $d$ DA gets mapped to a $1d$ HCLG.  For $1+1$ dimensional
  DAs in bulk in \cite{sumedha1} we have derived the exact expressions of
  $A(1,y)$ and $\Psi(x;1,y)$. Using them and
  Eq. \ref{density}-\ref{corr} we get the expressions for $A(w,y)$ and
  $\Psi(x;w,y)$ as follows

\begin{equation}
A(w,y) = \frac{2 w y (1+y)}{(1-y-wy-3wy^2)+(1+wy)\sqrt{(1-3y)(1+y)}}
\end{equation}

The connected density-density correlation function of the
corresponding gas is simple exponential and hence the generating
function $\Psi(x;w,y)$ has a form

\begin{equation}
\Psi(x;w,y) = K(w,y) \mbox{exp}(-{b(y)|x|}) \label{dcorr}
\end{equation}
where it is straightforward to calculate $K(w,y)$ and $b(y)$, and we
get

\begin{equation}
K(w,y) =
\frac{2wy(1-3y)(1+wy)(1-y+\sqrt{(1-3y)(1+y)})}{[(1-3y)(1+wy)\sqrt{1+y}+(1-y-wy-3wy^2)\sqrt{1-3y}]^2}
\end{equation}
and

\begin{equation}
b(y)=
\mbox{log}(\sqrt{1+y}+\sqrt{1-3y})-\mbox{log}(\sqrt{1+y}-\sqrt{1-3y})
\end{equation}

The generating functions $A(w,y)$ and $\Psi(x;w,y)$ have branch cut at
       $y=1/3$. For $w=1$, they also have a pole singularity at
       $y=1/3$. Hence, clearly the phase transition from desorbed to
       adsorbed phase occurs at $w=1$, i.e $w_c =1$. For $w \le 1$,
       $y_{\infty}(w) = 1/3$ and for $w > 1$ it is given by the real
       positive solution of

\begin{equation}
		   w-y-w (3+2 w) y^2 - 3 w^2 y^3=0
\end{equation}

Near the critical point, for $w=1+\delta$, to leading order we get
$y_{\infty}(w)$ to be
\begin{equation}
y_{\infty}(w) = \frac{1}{3} - \frac{{\delta}^2}{16}+ \mbox{higher order terms}
\end{equation}

The sticking fraction $C_{st}(w,y)$ can also be exactly calculated and we get it to be

\begin{equation}
C_{st}(w,y) = {\left[\frac{y(1-w)}{1+y}
+\frac{1+wy}{\sqrt{(1+y)(1-3y)}}\right]}^{-1}
\end{equation}

From this, near the critical point, we get the scaling form of
$C_{st}(w,y)$ to be

\begin{equation}
C_{st}(w,y) = \sqrt{\epsilon}~h(u) \label{2dscale}
\end{equation}
where $\epsilon  = 1-y/y_{\infty}(w)$ and $u = {\epsilon}^{-1/2}
\delta$ and we get
\begin{equation}
h(u) = \frac{\sqrt{3}}{2} \left[1+\frac{9 u^2}{48}\right]^{\frac{1}{2}}
\end{equation}

This gives the order parameter $C_{st}(w)$ near the critical point to
be proportional to $\frac{3 \delta}{8}$. For large values of $w$,
expanding $C_{st}(w)$ in powers of $(1/w)$ we get

\begin{equation}
     C_{st}(w) \sim \frac{1}{2} - \frac{3 \sqrt{2}}{16 \sqrt{w}}
-			\frac{3}{16 w}- .....
\end{equation}

 For $w \rightarrow \infty$ it approaches $1/2$, the maximum possible
	    fraction that can stick to wall, as expected.

\section{Two dimensional Directed Branched Polymer in presence of 1-d impenetrable surface}

In the presence of an impenetrable surface, because of loss in entropy
  per monomer on the wall, the transition from desorbed to adsorbed
  phase takes place at a non trivial value of adsorption activity.

\begin{figure}
\begin{center}
		 \epsfig{figure=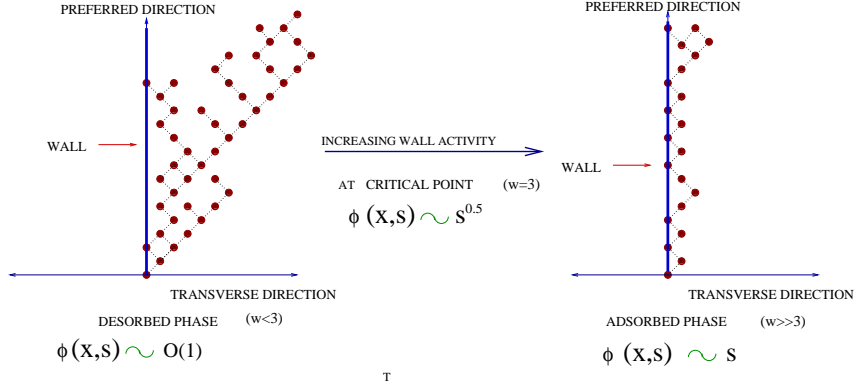,width=12cm}
\end{center}
\caption{Directed Branched Polymer on a Square lattice in presence of
a 1d impenetrable line about the diagonal.}
\end{figure}

    Here we study a DBP in $1+1$ dimension on a square lattice, in
  presence of an impenetrable surface, about the diagonal (Fig 2). From the exact generating function $A(1,y)$, $A(w,y)$ and
  $\Psi(x;w,y)$, it is straightforward to determine the critical value
  of $w$ and the sticking fraction and the density profile in the
  desorbed, critical and adsorbed phase of the system. The $1+1~d$ case in presence of a solid wall along the growth
  direction can be mapped to the HCLG in 1-d  with fugacity 0 for all
  sites lieing along the negative axis. Making use of this mapping we
  get

\begin{equation}
       A(1,y)= \frac{(1-y-\sqrt{1-2y-3y^2})}{2y}   \label{rens}
\end{equation}
and $A(w,y)$ is easy to get by substituting in Eq. \ref{density}. This
result can also be obtained using heap method. For an alternate
treatment see \cite{rensburg}.

The density-density correlation of the corresponding gas is
exponential and hence the generating function $\Psi(x;w,y)$ still has
a form given by Eq. \ref{dcorr}, but $K(w,y)$ and $b(y)$ are now given
by
\begin{equation}
K(w) = \frac{wy(1+wy)(\sqrt{1+y}+\sqrt{1-3y})}{(1+y) (1-w^2 y^2)
\sqrt{1-3y}+(1-y-(4-w) w y^2-w^2 y^3)\sqrt{1+y}}
\end{equation}
and
\begin{equation}
b(y) =
\mbox{log}(\sqrt{1+y}+\sqrt{1-3y})-\mbox{log}(\sqrt{1+y}-\sqrt{1-3y})
\end{equation}

The generating functions $A(w,y)$ and $\Psi(x;w,y)$ have a branch cut
     at $y=1/3$. At $w=1$, $A(1,y)$ has no divergence and
     $y_c=1/3$. Substituting in Eq. (\ref{pw}), we get $w_c=3$. This
     value is greater than the value for $1+1$ d DBP with a penetrable
     surface. This is expected, since the tendency of polymer to grow
     away from the surface is more when the surface is impenetrable
     and hence only when the surface gets sufficiently attractive, the
     polymer starts sticking to it. For $w>3$, the closest singularity
     to the origin occurs at

\begin{equation}
	    y_{s} = \frac{\sqrt{4w-3} -1}{2w} \label{sing}
\end{equation}

 For $w \leq 3$ the branch cut singularity $1/3$ dominates and hence
$y_{\infty}(w)$, the infinite polymer limit fugacity value is equal to
$1/3$ for $w \leq 3$.  Whereas for $w > 3$,
$y_{\infty}(w)=y_{s}$. Free energy is a constant and the order
parameter, $C_{st}(w)$ is zero for $w < 3$.

We get the  sticking fraction, $C_{st}(w,y)$ to be

\begin{equation}
	       C_{st}(w,y) = \frac{1-2y-3y^2+(-1+y+2
	       y^2)\sqrt{(1+y)(1-3y)}}{y[-2y+(w+2y-wy)\sqrt{(1+y)(1-3y)}+w(-1+2y+2y^2)]}
\end{equation}

Near the critical point for $w=3+\delta$ and $y=y_s (1-\epsilon)$, we
get the same scaling form for $C_{st}(w,y)$ as given by
Eq. \ref{2dscale}, with the scaling function $h(u)$ to be
\begin{equation}
h(u) = \frac{2}{\sqrt{3}} \left[1+\frac{u^2}{27}\right]^{\frac{1}{2}}
\end{equation}

Hence $C_{st}(w)$ is proportional to $\frac{2 \delta}{9}$ near the
  critical point and approaches $1/2$ as $w \rightarrow \infty$. This
  is plotted in Fig 3 along with the sticking fraction for the
  penetrable case. The qualitative behavior in both cases is just the
  same, the main difference being the shift of the transition point
  from $1$ to $3$ and the initial slope. For large value of $w$ it is
  easy to expand $C_{st}(w)$ in powers of $1/w$. It should be noted
  that large $w$ expansion of $C_{st}(w)$ will involve powers of
  $w^{-1/2}$ in this case as well.

\begin{figure}
\begin{center}
		 \epsfig{figure=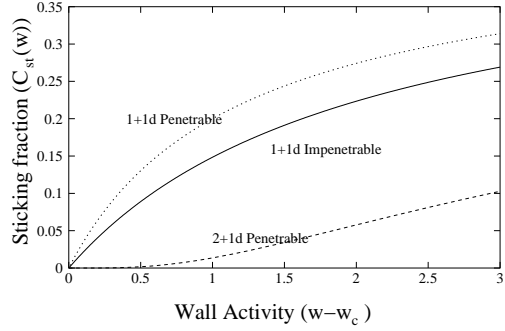,width=7cm}
\end{center}
     \caption{Sticking Fraction in presence of line for a Directed
     branched polymer in 1+1 and 2+1 dimensions, when the polymer size
     tends to infinity.}
\end{figure}

 Using the exact equations for generating function for $\Psi(x;w,y)$
    we translate these results to the constant number ensemble and we
    get the function $\phi(x,s)$ in three regions giving the spread of
    sites as a function of distance from the wall. Here we give these
    calculations for the impenetrable case only because the
    qualitative behavior in both impenetrable and penetrable case is
    exactly same for $1+1$ dimensional system.

 In the desorbed phase($w<3$), expanding near $y_c = 1/3$ as $y = y_c
   e^{-\epsilon}$, we get the scaling form for $\Psi(x;w,y)$ to be
\begin{equation}
\Psi(x;w,\epsilon ) =  c(w) \mbox{exp}(-x \sqrt{3 \epsilon})
\end{equation}
	   where, $c(w)$ is a $w$ dependent constant and is
 $\frac{3(3+w)}{2(3-w)^2}$.

     To obtain $\phi(x,s)$ for large $s$, we need to determine the
     coefficient of $y^s$ in the series expansion of
     $\Psi(x;w,\epsilon )$ i.e,

\begin{eqnarray}
\Psi(x;w,\epsilon ) \equiv \sum_{s} \phi(\textbf{x},{s}) A_s(w)~ y^s =
c(w) \sum_{k=0}^{\infty} \frac{(-\sqrt{3} x)^k}{\Gamma[k+1]} (1-3
y)^{\frac{k}{2}}\\ ~~~~~~~=c(w) \sum_{s=0}^{\infty} (3y)^s
\sum_{k=0}^{\infty} \frac{(-\sqrt{3}
x)^k}{\Gamma[k+1]}\frac{\Gamma[s-k/2]}{\Gamma[s+1] \Gamma[-k/2]}
\end{eqnarray}

For fixed $k$ and large $s$,

\begin{equation}
\frac{\Gamma[s-k/2]}{\Gamma[s+1]} \rightarrow s^{-1-k/2}
\end{equation}

Hence the leading singular behavior of $\phi(x,s)A_s(w)$ in the
desorbed phase is given by

\begin{equation}
\phi(x,s) A_s = \frac{3^s c(w)}{s}
\sum_{k=0}^{\infty}\frac{(-\sqrt{3}x/\sqrt{s})^k}{\Gamma[k+1]\Gamma[-k/2]}
\end{equation} 

Since $\Gamma[-k/2]$ has poles when $k$ is an even integer, only odd
terms contribute to the sum. It is easy to sum the resulting series,
giving $\phi(x,s)$ for large $s$ in the desorbed phase to be

\begin{equation}
\phi_{de}(x,s) = \frac{3}{2} x~ \mbox{exp}\left(-\frac{3x^2}{4s}\right)
\label{de}
\end{equation}

  For $w=3$, $c(w)$ is singular and we have to keep terms till first
  order in $\epsilon$ in the expansion(for $w< 3$ the constant term
  dominates) and we get

\begin{equation}
\Psi(x;3,\epsilon) = \frac{1}{\epsilon} \mbox{exp}(-x\sqrt{3\epsilon})
\end{equation}

 Again, just as in the desorbed phase expanding $\Psi(x;3,\epsilon)$
 in powers of $y^s$, the average number of sites at a distance $x$,
 i.e $\phi(x,s)$ for the critical region for large s is

\begin{equation}
		 \phi(x,s) = \frac{\sqrt{3\pi s}}{2}
   \mbox{erfc}\left(\frac{\sqrt{3}x}{2\sqrt{s}}\right) \label{cri}
\end{equation}

Hence we see that at $w=3$ not just the crossover exponent $\alpha$ is
   equal to $1/2$, but even the scaling form of $\phi(x,s)$ is same as
   that of a $(1+1)$ dimensional DA in bulk \cite{sumedha1} and hence
   same as that of the penetrable wall  at the critical point. This
   unusual result can be understood as coming from exact cancellation
   of decrease in entropy and increase in internal energy at the
   critical point. Also note that the value of exponent $\alpha =1/2$
   for DBP is equal to the estimates of $\alpha$ for branched polymers
   \cite{queiroz} and linear polymers \cite{guim} in $2$
   dimensions. Infact for adsorption of an undirected $d$ dimensional
   branched polymer to a $d-1$ dimensional surface, the crossover
   exponent $\alpha$ is conjectured to be $1/2$ in all spatial
   dimensions \cite{jenssen}.

 For $w > 3$, the behavior of the generating function is dominated by
  the singularity given by Eq. (\ref{sing}). For $w \gg 3$, $y_{s}
  \approx 1/\sqrt{w}$ and we get the large $s$ behavior of $\phi(x,s)$
  to be

\begin{equation}
	       \phi(x,s) = s~\mbox{exp}(-x) \label{ad}
\end{equation} 
  i.e, most of the sites stick to the origin as expected

Similarly, expanding $A(w,y)$ about $y_c$ and then going to constant
 number ($s$) ensemble, we get $A_s(w)$ for large $s$ as $A_s(w) \sim
 \frac{\sqrt{3}}{2\sqrt{\pi}} c(w)3^s s^{-\frac{3}{2}}$ in the
 desorbed regime. Hence the number of animals in presence of the $1d$
 impenetrable wall i.e $A_s(1)$ for large $s$ are $A_s(1)
 \sim\frac{\sqrt{3}}{2\sqrt{\pi}}3^s s^{-\frac{3}{2}}$. This gives
 ${\theta}_{de}$ to be $3/2$. This is consistent with the result
 derived for lattice trees by De'Bell et al \cite{debell}.  Also we
 get at the critical point $w=3$, $A_s(3) \sim
 \frac{2}{\sqrt{3\pi}}3^s s^{-\frac{1}{2}}$, implying ${\theta}_c$ to
 be $1/2$. For $w \gg 3$, $A_s(w) \sim (\sqrt{w})^s$, giving
 ${\theta}_{ad} =0$.

The function $\phi(x,s)$ gives the density profile of the polymer as a
 function of distance from the surface. Since the configurations are
 very different in two phases as shown schematically in Fig 2 hence
 $\phi(x,s)$ is very different in three regions. In desorbed phase  it
 peaks away from the surface at a distance of the order of the average
 transverse diameter of the polymer in the large $s$ limit. Whereas at
 the critical point it peaks at the surface (Fig 4).

\begin{figure}
\begin{center}
		\epsfig{figure=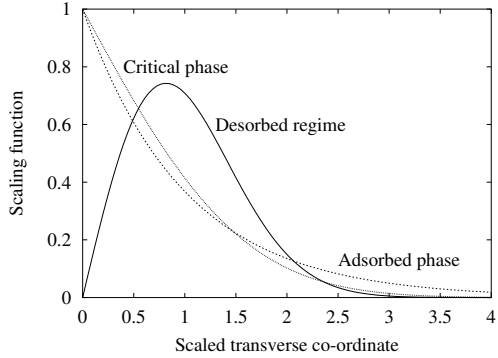,width=7cm}
\end{center}
   \caption{Denisty profile of a 2 dimensional DBP in presence of a
			     1dimensional surface }
\end{figure}

\section{Three dimensional Directed Branched Polymer in presence of  an attractive line}

    In $2+1$ dimensions, a DBP on a simple cubic lattice with nearest
and next nearest neighbor connections gets mapped to the hard hexagon
gas model in $2$ dimensions at negative activity in the disordered
regime, which was solved by Baxter \cite{baxter}. He obtained the
equation for the average density of the gas. It was shown by Joyce,
there is an algebraic equation in $z$ (activity of the gas) and $\rho$
(density of the gas) \cite{joyce1}.

   The equation given by Joyce is quartic in $z$ and 12th order in
  $\rho$. For convenience we will just rewrite it here \cite{joyce1}

\begin{equation}
    \rho (1-\rho)^{11} - (1-\rho)^5 P_1(\rho) z+ \rho^2 (1-\rho)^2
   P_2(\rho) z^2 - \rho^5 P_1(\rho) z^3+ \rho^{11} (1-\rho) z^4 = 0
\label{joy}
\end{equation}
where
\begin{eqnarray*}
   P_1(\rho) = (1-13 \rho +66 \rho^2 - 165 \rho^3 + 220 \rho^4 -165
 \rho^5+77 \rho^6-22 \rho^7)\\ P_2(\rho) = (1-13 \rho +63 \rho^2 -125
 \rho^3 +6 \rho^4 + 401 \rho^5 -689 \rho^6 + 476 \rho^7 - 119 \rho^8)
\end{eqnarray*}

The density $\rho$ of the HCLG is just the negative of $A(1,y)$ and $z
 =-y/(1+y)$. It is straightforward to get an algebraic equation in
 $A(1,y)$ as a function of $y$ \cite{joyce2}. As $A(w,y)$ is a simple
 rational function of $A(1,y)$, $y$ and $w$ (see Eq.\ref{density}),
 substituting $\rho$ in terms of $A(w,y)$, the grand partition
 function of the $2+1$ dimensional DA in presence of an one
 dimensional line about the main diagonal of the lattice, we get a
 $12$ th order polynomial equation in $A(w,y)$, where the coefficients
 are functions of $w$ and $y$. Explicit writing down the equation is
 rather tedious and is omitted. Since $A(1,y)$ becomes singular for $y
 = y_c = 2/(9+5\sqrt{5})$, hence in presence of 1-dimensional line the
 polymer will undergo a desorption-adsorption transition at $w=1$. For
 $w \leq 1$ the dominant singularity will be $y_c$ and $y_{\infty}(w)
 = y_c$. For $w >1$, at $y_{\infty}(w)$, $A(w,y)$ tends to infinity
 and at this point the coefficient of highest order term must be
 zero. Since we have a 12th order equation in $A(w,y)$ hence by
 equating the coefficient of the 12 th order term to zero, we get a
 polynomial equation in $y$ and $w$ ($Q(y,w)=0$) whose smallest
 positive real root would be $y_{\infty}(w)$. This polynomial is 12th
 order in $w$. But we can find the root numerically. The free-energy
 is just $log(y_{\infty}(w))$ and hence can be evaluated numerically.

    In this case the expressions of $A(w,y)$ and other generating
 functions are rather complicated and hence it is difficult to go to
 the constant size ensemble. But at critical point system behaves like
 bulk and since $\theta = 5/6$,  by hyper-scaling arguments $\nu_{c} =
 5/12$ which implies that the crossover exponent $\alpha = 1/6$
 (Eq. \ref{alp}). And by Eq. \ref{stscale}, the sticking fraction
 $C_{st}(w,y) \sim (1-y/y_{\infty}(w))^{5/6}$ as $w \rightarrow 1^+$
 asymptotically.

By solving $Q(y,w)=0$ we get $y_{\infty}(w)$ as a function of
$w$. Near the critical point for $w=1+\delta$, to leading order we get
\begin{equation}
y_{\infty}(w) = y_c(1-c \delta^6+\mbox{higher order term})
\end{equation}
where $c= 5 (5 \gamma)^5$ with $\gamma = (13 \sqrt{5}-25)/50$.

In the large polymer limit, for $y$ very close to $y_c$, $A(1,y)$ has
a scaling form

\begin{equation}
A(1,y) = a_0 \left(1-\frac{y}{y_c}\right)^{-\frac{1}{6}} \left[1+a_1
\left(1-\frac{y}{y_c}\right)^{\frac{5}{6}}+.....\right]
\end{equation}

where $a_0 = ({\sqrt{5}\gamma}^{1/6})^{-1}$ \cite{joyce2}.

Hence taking $y=y_{\infty}(w) (1-\epsilon)$ and $w=1+\delta$ we get
the scaling function of $C_{st}(w,y) = \epsilon^{5/6} h(u)$ to be
\begin{equation}
		      h(u) = \frac{6a_0}{1+y_c} (1+cu^6)^{\frac{5}{6}}
		      - 6 c u^5
\end{equation}
where $u = \delta \epsilon^{-1/6}$. The scaling function $h(u)$ is a
function of $w$ and $y$, which are both thermodynamic variables.

For large $w$, expanding in power of $1/w$ we get

\begin{equation}
  y_{\infty}(w) \sim \frac{1}{\sqrt{6w}} -\frac{1}{4w}-\frac{11}{16}
   \sqrt{\frac{3}{2}} \left({\frac{1}{w}}\right)^{\frac{3}{2}}-....
\end{equation}
and
\begin{equation}
	       C_{st}(w) \sim \frac{1}{2} - \frac{1}{4}
		 \sqrt{\frac{3}{2w}}-\frac{9}{4w}-...
\end{equation}

As $w \rightarrow \infty$, $C_{st}(w)$ approaches $1/2$, the maximum
  possible fraction of adsorption. It is like order parameter of the
  surface transition. It is plotted in Fig 3. As is clear from the
  scaling function, the sticking fraction increases much slowly than
  in the  $1+1$ $d$ case. This is expected as there the polymer in
  $d+1$ dimensions was getting adsorbed at a $d$ dimensional surface
  whereas here a polymer in $d+1$ dimensions is getting adsorbed on a
  $d-1$ dimensional surface.

I am very thankful to my advisor Prof. Deepak Dhar for guidance at
every step of this work. I acknowledge the partial financial support
received from the TIFR Alumni Association Scholarship.

\end{document}